\documentclass[aps,prx,reprint,nofootinbib,preprintnumbers,amsmath,amssymb,amsfonts]{revtex4-2}
\pdfoutput=1

\usepackage[utf8]{inputenc}
\usepackage{graphicx}
\usepackage{hyperref}

\def\be{\begin{equation}}
\def\ee{\end{equation}}

\begin{document}

\title{On the Field Excursion Bound}

\author{Tom Rudelius}
\email{thomas.w.rudelius@durham.ac.uk}
\affiliation{Department of Mathematical Sciences, Durham University, Durham DH1 3LE United Kingdom}

\date{\today}

\begin{abstract}
In a recent work \cite{Herderschee:2025nrf}, Herderschee and Wall (HW) proved a bound on scalar field excursions in spatially flat FRW cosmologies. In this note, we give an alternate proof of their bound using the Friedmann equations, and we prove that it can be saturated only in universes with vanishing acceleration, $\ddot a =0$. We argue that in a realistic (eternal) inflation scenario, the bound is robust against quantum corrections and spacetime curvature, and it can be violated by higher-derivative corrections only at the expense of a superluminal speed of sound. We further speculate on possible connections between the swampland program and the vacuum estimates given in the work of HW \cite{Herderschee:2025nrf}.
\end{abstract}


\maketitle

\section{Introduction}\label{sec:INTRO}

In \cite{Herderschee:2025nrf}, Herderschee and Wall proved a bound on scalar field excursions along null geodesics using the Raychaudhuri equation, which they named the Field Excursion Bound (FEB).

In an FRW cosmology with vanishing spatial curvature ($k=0$), the FEB amounts to the bound
\be
\sqrt{2} \kappa_4 |\Delta \phi| \leq \Delta N + \log\left( \frac{H(t_i)}{H(t_f)} \right)\,,
\label{FLEB}
\ee
where $\kappa_4^2 = 8 \pi G = M_{\rm Pl}^{-2}$, $H(t) = \dot a/a$ is the Hubble parameter, $|\Delta \phi| = |\phi(t_f) - \phi(t_i)|$ is the length of the scalar field excursion from time $t_i$ to time $t_f$, and $\Delta N= \log(a(t_f)/a(t_i))$ is the number of e-folds of cosmic expansion between $t_i$ and $t_f$. In what follows, we will refer to \eqref{FLEB} as the FLat Excursion Bound (FLEB).\footnote{We thank Aron Wall for suggesting this name to us.} This equation will be the primary focus of this paper.

Our main results are as follows. In \S\ref{sec:SIMPLE}, we show that the FLEB follows simply from the Friedmann equations. As a corollary, it can be saturated only when $\ddot a = 0$ and all other sources of stress-energy saturate the NEC, $\rho_i + p_i = 0$. In \S\ref{sec:QUANTUM}, we show that the FLEB is satisfied comfortably during slow-roll inflation, and we demonstrate quantitatively that quantum corrections are unlikely to violate the bound during inflation. We also derive a sort of converse to the FLEB. In \S\ref{sec:DERIVATIVE}, we show that the FLEB can be violated by higher-derivative corrections; however, for the class of models we consider, such a violation requires a superluminal speed of sound. In \S\ref{sec:CURVATURE}, we show that the FLEB is satisfied in open universes, while it may be violated in closed universes.
We conclude in \S\ref{sec:SWAMPLAND} by speculating on possible connections between the vacuum estimates used \cite{Herderschee:2025nrf} and the swampland program.


\section{Simple Derivation of the Flat Excursion Bound}\label{sec:SIMPLE}

In this section, we show that \eqref{FLEB} follows simply from the Friedmann equations, and we prove that it can be saturated only when $\ddot a = 0$ and all other sources of stress energy have $w_i = p_i/\rho_i = -1$.

We begin with the action
\be
S = \int \sqrt{-g} \left( \frac{1}{2 \kappa_4^2} R - \frac{1}{2} \partial_\mu \phi \partial^\mu \phi - V((\phi)\right)\,.
\ee
Here and throughout this paper, we use a $(-, +, +, +)$ metric signature. We further assume spatial homogeneity for the scalar field, $\phi(t,x) = \phi(t)$, in which case the energy density and pressure take the familiar form
\be
\rho = \frac{1}{2} \dot \phi^2 + V(\phi) \,,~~~ p = \frac{1}{2} \dot \phi^2 - V(\phi)\,,
\ee
so
\be
\rho + p = \dot \phi^2\,.
\ee
The Friedmann equations for a spatially flat ($k=0$) universe further give 
\be
\rho + p = -\frac{2}{\kappa_4^2} \dot H \,.
\ee

With these formulae in hand, we begin our proof of the FLEB \eqref{FLEB} with a trivial inequality:
\be
0 \leq \frac{1}{H^2}(\dot H + H^2)^2\,.
\label{triv}
\ee
This inequality is saturated if and only if $\dot H = - H^2$, which is equivalent to the condition $\ddot a = 0$.

Next, we note that \eqref{triv} can be rewritten as
\be
-4 \dot H \leq H^2 - 2 \dot H + \frac{\dot H^2}{H^2}\,.
\ee
Setting $-2 \dot H  = \kappa_4^2 (\rho+p ) = \kappa_4^2 \dot \phi^2$ and using the relation $\dot N = H$, this gives
\be
2\kappa_4^2 \dot \phi^2 \leq (\dot N - \frac{\dot H}{H})^2\,.
\label{FLEBderivsquared}
\ee
Finally, we take the square root to give\footnote{Note that both terms on the right-hand side of \eqref{FLEBderiv} are non-negative in an expanding universe that satisfies the NEC, justifying the sign of the square root.}
\be
\sqrt{2}\kappa_4 |\dot \phi| \leq \dot N - \frac{\dot H}{H}\,.
\label{FLEBderiv}
\ee
This equation can be viewed as a derivative form of the FLEB. Integrating, we recover the original FLEB, \eqref{FLEB}.
Furthermore, from \eqref{triv}, we see that saturation of the FLEB occurs if and only if $\ddot{a}(t) = 0$ for all $t \in [t_i, t_f]$.
Note that this condition further implies that the two terms on the right-hand side of \eqref{FLEB} are equal, $\Delta N = \log\left( \frac{H(t_i)}{H(t_f)} \right)$.

This result generalizes to cases with other sources of stress-energy that satisfy the NEC. In such cases, we may write
\be
\rho + p = \dot \phi^2 + \sum_i (1+w_i) \rho_i\,,
\ee
where $w_i \geq -1$, $\rho_i \geq 0$. From above, we have that 
\be
2 (\rho + p) \leq (\dot N - \frac{\dot H}{H})^2
\,,
\ee
with saturation occurring only when $\ddot a = 0$. Furthermore,
\be
\dot \phi^2 = \rho + p -  \sum_i (1+w_i) \rho_i \leq \rho + p \,,
\ee
with saturation occurring only when $\sum_i (1+w_i) \rho_i = 0$. We conclude that the derivative version of the FLEB \eqref{FLEBderiv} can be saturated only when $\ddot a= 0$ and all other sources of nonzero energy density $\rho_i$ have $w_i =-1$. Integrating from time $t=t_i$ to $t=t_f$, the same result applies to the original version of the FLEB \eqref{FLEB}.\footnote{For similar bounds on scalar field traversals in FRW cosmologies with matter, see \cite{Andriot:2024sif}.}

This result generalizes straightforwardly to theories with multiple scalar fields and nontrivial moduli space metrics $G_{AB}(\phi)$, with $\dot \phi^2$ replaced by $G_{AB} \dot \phi^A \dot \phi^B$; here, $|\Delta \phi|$ in \eqref{FLEB} should be understood as the distance along the path $\phi(t)$ from $t =t_i$ to $t=t_f$ as measured by $G_{AB}(\phi)$.


\section{Quantum Corrections in the Slow-Roll Regime}\label{sec:QUANTUM}

The fact that the FLEB can be saturated only when $\ddot a = 0$ suggests that it will be satisfied comfortably during inflation, where $\ddot a$ grows exponentially with time. Indeed, this expectation may be quantified using the first slow-roll parameter:
\be
\varepsilon \equiv - \frac{\dot H}{H^2}\,.
\ee
The condition for saturation $\ddot a = 0$ is then equivalent to the condition $\varepsilon = 1$. 
More generally, we wish to quantify the distance from saturation. To this end, we define
\be
\Delta_F \equiv (\dot N - \frac{\dot H}{H})^2 - 2 \kappa_4^2 \dot \phi^2 \,,
\label{DeltaF}
\ee
which is the difference between the left-hand side and right-hand side of  \eqref{FLEBderivsquared}. Saturation of the FLEB is equivalent to the condition $\Delta_F = 0$, and a violation of the FLEB would require $\Delta_F < 0$.

We may then write
\be
\dot \phi^2 = \rho + p = -\frac{2}{\kappa_4^2} \dot H = \frac{2}{\kappa_4^2} \varepsilon H^2\,,
\ee
which implies
\be
\Delta_F = (1 - \varepsilon)^2 H^2\,,
\label{DeltaFresult}
\ee
hence $\Delta_F \approx H^2$ in the slow-roll regime $\varepsilon \ll 1$. This quantifies the distance from saturation of the FLEB during slow-roll inflation.

During inflation, quantum fluctuations exit the horizon and behave classically \cite{Kiefer:1998qe}. These fluctuations are Gaussian (to leading order), with typical size \cite{Vilenkin:1982wt, Linde:1982uu, Starobinsky:1982ee}
\be
\delta \phi_q  = \frac{H}{2 \pi}\,,
\ee
per Hubble time $\delta t_H = H^{-1}$. This means that we may expect quantum corrections to $\dot \phi$ of order $\delta \phi_q/\delta t_H 
\sim H^2/(2 \pi)$. These induce corrections to $\Delta_F$ of the form
\begin{align}
\begin{split}
\Delta_{F,q} - \Delta_F &=  - 2 \kappa_4^2 \left( (\dot \phi + \frac{\delta \phi_q}{\delta t_H})^2 - \dot \phi^2 \right)  \\
&\sim - \frac{2 \sqrt{2 \varepsilon}}{\pi} \kappa_4 H^3  - \frac{\kappa_4^2}{2 \pi^2}  H^4 \,,
\end{split}
\end{align}
which are suppressed relative to the classical result $\Delta_F \approx H^2$ by factors of $\kappa_4 H = H/M_{\rm Pl}$ and $\sqrt{\varepsilon}$. We conclude that in the controlled slow-roll regime with $\varepsilon \ll 1$ and $H/M_{\rm Pl} \ll 1$, quantum corrections are exponentially unlikely to violate the FLEB.

This quantitative result agrees with the qualitative expectation set forth in \cite{Herderschee:2025nrf}.

To conclude this section, we remark in passing that \eqref{DeltaFresult} also allows us to prove a sort of converse to the FLEB. Combining with \eqref{DeltaF} and invoking the triangle inequality, we have
\be
\sqrt{2}\kappa_4 |\dot \phi| + |1 - \varepsilon| H \geq \dot N - \frac{\dot H}{H} 
\ee
Using the relations $\dot N =H$, $\varepsilon = - \dot H/H^2$, this yields
\be
\sqrt{2}\kappa_4 |\dot \phi|  \geq \begin{cases}
2 H & \varepsilon \geq 1 \\
- 2\frac{\dot H}{H} & \varepsilon < 1 
\end{cases} \,.
\ee
Integrating, we find
\be
\sqrt{2} \kappa_4 |\Delta \phi| \geq   2 \cdot   \text{min}\left(\Delta N ,\log\left( \frac{H(t_i)}{H(t_f)} \right)\right) \,. \label{conversebound}
\ee
In other words, the left-hand side of the FLEB \eqref{FLEB} is at least twice as large as the smaller of the two terms on the right-hand side. This bound is also saturated if and only if $\ddot a(t) = 0$ identically.

In a slow-roll inflation scenario with $\varepsilon < 1$, we have $\log(\frac{H(t_i)}{H(t_f)}) < \Delta N$, so the second argument on the right-hand side of \eqref{conversebound} is the relevant one. Morally speaking, this bound is weaker than the Lyth bound \cite{Lyth:1996im} by a factor of $\sqrt{\varepsilon}$, so it does not offer a phenomenologically relevant bound on the tensor-to-scalar ratio.


\section{Higher-derivative Corrections}
\label{sec:DERIVATIVE}

The derivations of the FLEB in \cite{Herderschee:2025nrf} and \S\ref{sec:SIMPLE} apply for an arbitrarily number of scalar fields with an arbitrarily moduli space metric. However, those derivations assume a two-derivative action. In this section, we consider corrections from higher-derivative terms, such as those appearing in the DBI action \cite{Silverstein:2003hf}.

Following \cite{Garriga:1999vw, Armendariz-Picon:2000ulo}, we consider the ``$k$-essence'' scalar field action:
\be
S_{\phi} = \int d^4x \sqrt{-g} [P(X, \phi) - V(\phi)]\,,
\label{Sgen}
\ee
where $X \equiv - \frac{1}{2} g^{\mu \nu} \partial_\mu \phi \partial_\nu \phi$. Examples of such actions include the vanilla case of a canonically normalized scalar field, which has $P_{\rm cnc}(X, \phi) = X$, and the DBI action for a D3-brane in a warped throat region, which has $P_{\rm DBI}(X, \phi) = \frac{2\phi^4}{\lambda}\sqrt{1 - \frac{\lambda X}{\phi^4}}$ \cite{Silverstein:2003hf}. We will again restrict our attention to the case of a homogenous scalar field $\phi(t, x) = \phi(t)$, in which case $X = \frac{1}{2} \dot \phi^2$.

Varying \eqref{Sgen} with respect to the metric, one obtains the energy density and pressure for the scalar field condensate in a spatially flat FRW cosmology:
\begin{equation}
\begin{split}
\rho &= 2 X P_{,X}(X, \phi) - P(X, \phi) + V(\phi)  \\
p &= P(X, \phi) - V(\phi)\,,
\end{split}
\end{equation}
where $P_{,X} = \partial_X P$.
Thus, the Friedmann equations imply
\be
\rho + p = 2 X P_{,X} = -\frac{2}{\kappa_4^2} \dot H\,.
\label{rhopluspHD}
\ee
From here, we may follow the same steps as in \S\ref{sec:SIMPLE} above to derive a generalized version of the derivative form of the FLEB:
\be
2 \kappa_4 \sqrt{X P_{,X}} \leq \dot N - \frac{\dot H}{H}\,.
\ee
Setting $X = \frac{1}{2} \dot \phi^2$ and integrating, we find a higher-derivative version of the FLEB:
\be
\sqrt{2} \kappa_4 \int_{\phi_i}^{\phi_f} d \phi \sqrt{P_{,X}} \leq \Delta N + \log\left( \frac{H(t_i)}{H(t_f)} \right)\,.
\label{HDFLEB}
\ee
Note that this bound indeed reduces to the ordinary FLEB for the case of a canonically normalized scalar field, $P_{\rm cnc} = X$. Furthermore, while our derivation above was specific to the context of a flat FRW cosmology, a more general FEB for theories with higher-derivative kinetic terms can be obtained by following the analysis of \cite{Herderschee:2025nrf}; this is shown in Appendix \ref{App}.

Let us now consider the question: under what conditions can higher-derivative corrections invalidate the original FLEB? From \eqref{HDFLEB}, we see that a violation of the FLEB requires $P_{,X} < 1$, since if $P_{, X} \geq 1$, then
\be
|\Delta \phi| \leq \int_{\phi_i}^{\phi_f} d \phi \sqrt{P_{,X}}\,,
\ee
in which case the FLEB \eqref{FLEB} follows immediately from \eqref{HDFLEB}. One may easily check that the D3-brane DBI Lagrangian $P_{\rm DBI}(X, \phi) = \frac{2\phi^4}{\lambda}\sqrt{1 - \frac{\lambda X}{\phi^4}}$ satisfies $P_{,X}(X, \phi) \geq 1$ for all $X \geq 0$, so the FLEB holds even in DBI inflation.

More generally, a scalar field with action given by \eqref{Sgen} has a ``speed of sound'' $c_s$ given by
\cite{Garriga:1999vw, Chen:2006nt}
\be
c_s^2 = \frac{P_{,X}}{P_{,X} + 2 X P_{,XX}}\,.
\label{soundspeed}
\ee
Note that $X$, $P_{, X}$ are both non-negative for a homogenous scalar field that satisfies the NEC.

Let us normalize our scalar field so that $P_{,X}(X = 0, \phi) = 1$; this ensures that the scalar field is canonically normalized at low energies, where higher-derivative corrections are irrelevant. We then have
\begin{align}
P_{,X}(X, \phi) &= P_{,X}(X=0, \phi) + \int_{X'=0}^X dX' P_{,XX}(X', \phi) \nonumber \\
&= 1 + \int_{X'=0}^X dX' P_{,XX}(X', \phi)\,.
\end{align}
From this, we see that the necessary condition to violate the FLEB, $P_{,X}(X, \phi) < 1$, requires $P_{,XX}(X', \phi) < 0$ for some $X'$. From \eqref{soundspeed}, we see that $P_{,XX} < 0$ in turn implies $c_s > 1$. In other words, a violation of the FLEB via higher-derivative corrections requires a superluminal speed of sound.

It has been argued that such superluminal propagation of scalar field perturbations in FRW backgrounds does not preclude a self-consistent notion of causality \cite{Bruneton:2006gf, Kang:2007vs, Armendariz-Picon:2005oog, Babichev:2007dw, Bessada:2009ns}. However, as shown in \cite{Adams:2006sv}, there exist other backgrounds for which such superluminal propagation \emph{does} lead to a breakdown of causality. This suggests that theories with $c_s > 1$ likely do not admit a UV completion in quantum gravity (i.e., they likely reside in the swampland).


\section{Spacetime Curvature}\label{sec:CURVATURE}

Let us next consider what happens when we introduce spatial curvature. The FRW metric takes the form
\be
ds^2 = -dt^2 + a(t)^2 \left[\frac{dr^2}{1 - k r^2} + r^2 (d \theta^2 + \sin^2 \theta d \phi^2) \right]\,.
\ee
The Friedmann equations now give
\be
\frac{1}{\kappa_4^2} \dot H = \frac{k}{a^2} - \frac{1}{2}(\rho + p)\,.
\ee
Hence,
\be
\kappa_4^2 \dot \phi^2 - \frac{2 k}{a^2} = \rho + p  - \frac{2 k}{a^2} = - 2 \dot H   \,.
\ee
Following the steps of \S\ref{sec:SIMPLE}, we thus find the derivative version of the FLEB \eqref{FLEBderiv}, generalized to the case of nonzero spacetime curvature:
\be
\sqrt{2} \left( \kappa_4^2 \dot \phi^2  - \frac{2k}{a^2} \right)^{1/2} \leq \dot N - \frac{\dot H}{H}\,.
\label{kFEBderiv}
\ee
For $k \leq 0$, we have 
\be
\kappa_4 |\dot \phi| \leq  \left( \kappa_4^2 \dot \phi^2  - \frac{2k}{a^2} \right)^{1/2}\,,
\ee
which (upon integrating) implies the FLEB. Thus, the FLEB is automatically satisfied in both flat ($k=0$) and open ($k<0$) universes. It may, in principle, be violated in closed universes ($k>0$), though a modified version of it may be obtained by integrating both sides of \eqref{kFEBderiv}:
\be
\sqrt{2} \int_{t_i}^{t_f}dt  \left( \kappa_4^2 \dot \phi^2  - \frac{2k}{a^2} \right)^{1/2} \leq \Delta N + \log\left( \frac{H(t_i)}{H(t_f)} \right)\,.
\label{kFEB}
\ee

In an inflationary regime, $k/(a H)^2$ tends to zero at late times, in which case the curvature correction in \eqref{kFEB} is negligible. Furthermore, bubble universes created by Coleman-de Luccia tunneling are negatively curved ($k < 0$) \cite{Coleman:1980aw}. This all suggests that the FLEB will be satisfied comfortably in a realistic (eternal) inflation scenario.


\section{Swampland Bounds on Vacua}\label{sec:SWAMPLAND}

As one possible application of the FLEB, \cite{Herderschee:2025nrf} argued that the number of e-folds of inflation required for anthropic vacuum selection is likely to be much larger than the $\approx 60$ e-folds required by observation. In this section, we give a lightning review of their argument, and we explore connections between this argument and recent ideas in the swampland program. We stress that this section is relatively speculative in nature, hence we conclude it with a discussion of key assumptions and possible points of failure.

To begin, \cite{Herderschee:2025nrf} crudely estimated that explaining the $10^{-120}$ tuning of the cosmological constant and the $10^{-30}$ tuning of the Higgs mass requires at least $\sim 10^{150}$ vacua.

Taylor expanding the scalar potential $V(\phi^A)$ and ignoring irrelevant terms of degree $d \geq 5$, the condition for a vacuum $\partial_A V = 0$  amounts to a system of $n_s$ cubic equations, where $n_s$ is the number of scalar fields. By B\'ezout's theorem, this has $3^{n_s}$ solutions over the complex numbers. Many of these solutions are likely to be complex and/or unstable critical points, but nonetheless $3^{n_s}$ represents an upper bound on the number of vacua in the region of moduli space where the Taylor expansion holds. We might reasonably take this region to be an $n_s$-dimensional ball of radius $M_{\rm Pl}$.

Next, \cite{Herderschee:2025nrf} bounded the total number of vacua $n_v$ inside a region of moduli space of volume $V$ as
\be
n_v \lesssim 3^{n_s} \cdot \frac{V}{\text{vol}(B_{n_s}(M_{\rm Pl}))}\,,
\ee
where $\text{vol}(B_{n_s}(M_{\rm Pl}))$ is the volume of an $n_s$-dimensional ball of radius $M_{\rm Pl}$.
If the moduli space is flat, then this estimate yields $n_v \sim (3|\Delta \phi|)^{n_s}$, and attaining the desired number of vacua $n_v \gtrsim 10^{150}$ requires $|\Delta \phi| \gg M_{\rm Pl}$ for $n_s \lesssim 100$. Via the FLEB \eqref{FLEB}, this in turn requires $\Delta N \gg 1$.

The authors of \cite{Herderschee:2025nrf} pointed out that this conclusion could be evaded if the moduli space is negatively curved, in which case the estimate $n_v \sim (3|\Delta \phi|)^{n_s}$ is no longer valid, and instead the accessible volume of moduli space inside a region of size $|\Delta \phi|$ grows exponentially with $|\Delta \phi|$. However, this apparent loophole may perhaps be closed by the arguments of \cite{Delgado:2024skw}, which reasoned that the finiteness of amplitudes in quantum gravity (after compactifying to one dimension) requires the moduli space volume of a ball of radius $|\Delta \phi|$ to grow no faster than $|\Delta \phi|^{n_s+ \epsilon}$ for arbitrarily small $\epsilon >0$. The derivation of this result given in \cite{Delgado:2024skw} relied on supersymmetry, as it assumed an exact moduli space ($V(\phi^A) = 0$). Nonetheless, the result may apply more generally, particularly in string compactifications where the space of (massive) scalar fields is inherited from an underlying supersymmetric moduli space of massless scalar fields.

Another relevant insight from the last decade of swampland-related research is that the fertile ground for (de Sitter) vacua in quantum gravity seems unlikely to span parametrically super-Planckian distances in scalar field space. It is well-established that the existence of a vacuum requires competition of multiple terms in a perturbative/non-perturbative expansion \cite{Kachru:2003aw, Balasubramanian:2005zx}, which cannot occur arbitrarily far out in scalar field space. Instead, at very large scalar field vevs $||\phi^A|| \gg M_{\rm Pl}$, the potential $V(\phi^A)$ decays exponentially as $V \sim \exp(-c  ||\phi^A||)$ \cite{Obied:2018sgi, Bedroya:2019snp, Rudelius:2022gbz}. Likewise, a least one tower of massive particles has a characteristic mass scale $m$ that decays exponentially as $m \sim \exp(-\alpha ||\phi^A||)$ \cite{Ooguri:2006in}.

In \cite{Baume:2016psm}, Baume and Palti argued on the basis of simple examples that the exponential decay behavior of particle masses should kick in within an $O(M_{\rm Pl})$ distance. This perhaps suggests that the asymptotic regime of scalar field space, where the potential itself decays exponentially, is no more than an $O(M_{\rm Pl})$ distance away from any point in field space \cite{Scalisi:2018eaz, vandeHeisteeg:2023uxj}. If so, then the number of vacua in a ball of radius $|\Delta \phi|$ may be estimated as
\be
n_v \lesssim \begin{cases} 
     (3 |\Delta \phi|/M_{\rm Pl})^{n_s} & |\Delta \phi | \lesssim M_{\rm Pl} \\
      (3 k)^{n_s} & |\Delta \phi | \gg M_{\rm Pl}\,,
   \end{cases}
   \label{nvest}
\ee
where $k$ is some order-one fudge factor that corresponds to the radius (in Planck units) of the region in moduli space where vacua can be found.\footnote{Here, ``order-one'' should perhaps be taken with a grain of salt. The original example of LVS moduli stabilization \cite{Balasubramanian:2005zx} yields a vacuum at a proper field distance $|\Delta \phi| \approx 20 M_{\rm Pl}$ from the tip of the stretched K\"ahler cone, as defined in \cite{Demirtas:2018akl}. We thank Matthew Reece for discussions on this point.}
In particular, for $n_s \lesssim 100$, we likely have $n_s \ll 10^{150}$. Thus, for a typical choice of fluxes, there will be zero anthropically viable vacua across scalar field space.

In summary, the picture emerging from the swampland program suggests that vacua are relatively sparse in scalar field moduli space: for a given choice of fluxes, the number of vacua is not exponentially large, and the vacua that do exist are bunched together within a distance of $O(M_{\rm Pl})$. If this picture is correct, then encountering a viable vacuum likely requires not only a large scalar field excursion, but also a flux-changing phase transition \cite{Brown:1988kg}.

Once again, it should be emphasized that the reasoning of this section is rather speculative. The aforementioned work of \cite{Delgado:2024skw} and \cite{Baume:2016psm} relied on the presence of supersymmetry, and at present it is unclear whether their results should carry over to more realistic, non-supersymmetric landscapes. In addition, the work of \cite{Baume:2016psm} focused specifically on the behavior of towers of light particles, so the application of their results to scalar field potentials requires a further degree of extrapolation; progress in this direction can be found in \cite{ vandeHeisteeg:2023uxj}. Finally, although the spirit of \cite{Baume:2016psm} has been generally affirmed by subsequent studies, the precise version of their proposed bound remains a matter of ongoing discussion and debate \cite{vandeHeisteeg:2023uxj, Scalisi:2018eaz, Hebecker:2017lxm,   Rudelius:2023mjy}. 

Conversely, it is worth noting that our estimates above equated critical points of the potential with stable vacua. With increasing $n_s$, this approximation becomes increasingly suspect, and hence it is very possible that \eqref{nvest} badly overestimates the number of vacua for the given configuration of fluxes. Regardless, it is intriguing that recent progress in the swampland program may point to upper bounds on the number of vacua for a given configuration of fluxes. Further research in this direction is needed.


\vspace*{.5cm}
{\bf Acknowledgments.} We thank Matthew Reece and Simon Ross for useful discussions. We thank Aidan Herderschee, Matthew Reece, Marco Scalisi, and Aron Wall for comments on a draft.
This work was supported by the STFC through grant ST/T000708/1.
\vspace*{.5cm}

\appendix

\section{FEB for General Kinetic Terms}\label{App}

In this appendix, we present an alternate derivation of \eqref{HDFLEB}, which parallels the treatment in \cite{Herderschee:2025nrf} for the case of a canonically normalized scalar field. We begin with the bound
\be
\frac{d \theta}{d \lambda} \leq - \frac{\theta^2}{2} - R_{\lambda\lambda} \,,
\label{Rayeq}
\ee
which follows from the Raychaudhuri equation \cite{Raychaudhuri:1953yv, Sachs:1961zz}. Here, $\theta$ is the expansion parameter and $\lambda$ is the affine parameter for a family of null geodesics. Following \cite{Herderschee:2025nrf}, we consider a null congruence with null tangent vector $k^\mu$ and auxiliary null vector $l^\mu$ satisfying
\be
k^\mu \nabla_\mu k^\nu = 0\,,~~~k \cdot l = -1\,,~~~k^\mu \nabla_\mu l^\nu = 0\,.
\ee
With this, we have
\be
R_{\lambda \lambda} = k^\mu k^\nu R_{\mu \nu} = \kappa_4^2 k^\mu k^\nu T_{\mu\nu}= \kappa_4^2 \left(\frac{dt}{d\lambda}\right)^2 (\rho + p)\,,
\ee
where we used the Einstein equations in the second equality and the FRW stress-energy tensor in the third equality. Invoking \eqref{rhopluspHD} and assuming a homogenous scalar field, we may further write
\be
\rho + p = 2 X P_{, X} = \dot \phi^2 P_{, X}\,,
\ee
so
\be
R_{\lambda \lambda} = \kappa_4^2 \left(\frac{d\phi}{d\lambda}\right)^2 P_{, X}\,.
\ee
Plugging this into
\eqref{Rayeq} and dividing both sides by $y \equiv  (d \phi/d \lambda) \sqrt{P_{,X}}$, we find
\be
\left| \frac{1}{\sqrt{P_{,X}}} \frac{d \theta}{d \lambda} \left(\frac{d \phi}{d \lambda}\right)^{-1} \right| \geq \left| \frac{\theta^2}{2 y}  + y \kappa_4^2  \right| \,.
\ee
The right-hand side can be minimized to give
\be
y^2 = \frac{\theta^2}{2 \kappa_4^2}\,.
\ee
Hence,
\be
\left| \frac{1}{\sqrt{P_{,X}}} \frac{d \theta}{d \phi}\right| \geq |\sqrt{2}\kappa_4 \theta|\,.
\ee
Integrating, we find
\be
\sqrt{2} \kappa_4 \int_{\phi_i}^{\phi_f} d \phi \sqrt{P_{,X}} \leq \log(\frac{\theta_f}{\theta_i})\,.
\ee
Finally, using the flat FRW result $\theta  = H/a$, we recover \eqref{HDFLEB}.

\bibliography{ref}

\end{document}